\documentclass[11pt]{article}
\usepackage{amssymb}
\usepackage{amsmath}
\usepackage[all]{xy}
\usepackage{graphicx}

\title{{\sc Beyond Purity and Mixtures in\\Categorical Quantum Mechanics}}

\author{{\sc Christian de Ronde}$^{1,2,3}$ and {\sc C\'esar Massri}$^{4,5}$}
\date{}

\usepackage[margin=2 cm]{geometry}
\newcommand{\fig}{\linebreak\refstepcounter{figure}{\small Fig. \thefigure}}

\begin{document}

\bibliographystyle{plain}
\maketitle

\begin{center}
\begin{small}
1. Philosophy Institute Dr. A. Korn, University of Buenos Aires - CONICET\\
2. Center Leo Apostel for Interdisciplinary Studies\\Foundations of the Exact Sciences - Vrije Universiteit Brussel\\
3. Institute of Engineering - National University Arturo Jauretche\\
4. Institute of Mathematical Investigations Luis A. Santal\'o, UBA - CONICET\\
5. University CAECE
\end{small}
\end{center}

\begin{abstract}
\noindent In a recent paper \cite{deRondeMassri19a}, we discussed the serious inconsistency present within the operational and mathematical definition(s) of the notion of {\it pure state}. Continuing this analysis, in this work we attempt to address the role of `purity' and `mixtures' within two different categorical approaches to QM, namely,  the topos approach originally presented by Chris Isham and Jeremy Butterfield \cite{Isham97, IshamButter1, IshamButter2} and the more recent logos categorical approach presented by the authors of this article \cite{deRondeMassri18a, deRondeMassri18b, deRondeMassri19b}. While the first approach exposes the difficulties to produce a consistent understanding of pure states and mixtures, the latter approach presents a new scheme in which their reference is erased right from the start in favor of an intensive understanding of projection operators and quantum superpositions. This new account of the theory, grounded on an intensive interpretation of the Born rule, allows us not only to avoid the orthodox interpretation of projection operators ---either as referring to definite valued properties or measurement outcomes--- but also to consider all matrices (of any rank) on equal footing. It is from this latter standpoint that we conclude that instead of distinguishing between pure and mixed states it would be recommendable ---for a proper understanding of the theory of quanta--- to return to the original matrix formulation presented by Werner Heisenberg in 1925. 
\end{abstract}
\begin{small}

\end{small}

\newtheorem{theo}{Theorem}[section]
\newtheorem{definition}[theo]{Definition}
\newtheorem{lem}[theo]{Lemma}
\newtheorem{met}[theo]{Method}
\newtheorem{prop}[theo]{Proposition}
\newtheorem{coro}[theo]{Corollary}
\newtheorem{exam}[theo]{Example}
\newtheorem{rema}[theo]{Remark}{\hspace*{4mm}}
\newtheorem{example}[theo]{Example}
\newcommand{\proof}{\noindent {\em Proof:\/}{\hspace*{4mm}}}
\newcommand{\qed}{\hfill$\Box$}
\newcommand{\ninv}{\mathord{\sim}} 
\newtheorem{postulate}[theo]{Postulate}

\bigskip

\bigskip

\bigskip

\section{Pure States and Mixtures in Textbook QM}

Today, the orthodox textbook formulation of standard Quantum Mechanics (QM)  taught in all universities around the world makes use of the notion of {\it pure state}. This notion, in turn, plays an essential role within the many debates taking place within the foundational and philosophical literature about QM and Quantum Information. Although historically the notion of pure state began to be mentioned, in the context of QM, during the late 1940s ---specially in relation to statistical theories---, the main ideas behind it were already presented by Paul Dirac in his famous book, {\it The Principles of Quantum Mechanics} \cite{Dirac74}, published in 1930. Following the positivist understanding of physical theories, Dirac argued that it is ``important to remember that science is concerned only with observable things and that we can observe an object only by letting it interact with some outside influence. An act of observation is thus necessarily accompanied by some disturbance of the object observed.'' In line with the positivist stance, Dirac also stressed the superfluous role of (metaphysical) representations: ``it might be remarked that the main object of physical science is not the provision of pictures, but the formulation of laws governing phenomena and the application of these laws to the discovery of phenomena. If a picture exists, so much the better; but whether a picture exists of not is a matter of only secondary importance.'' Dirac was then forced to address the sudden appearance of measurement outcomes from quantum superpositions, a notion that was regarded by him as kernel to the new theory: ``The nature of the relationships which the superposition principle requires to exist between the states of any system is of a kind that cannot be explained in terms of familiar physical concepts. One cannot in the classical sense picture a system being partly in each of two states and see the equivalence of this to the system being completely in some other state.'' In order to solve the positivist dilemma of quantum observability Dirac introduced a ``quantum jump''\footnote{A notion that Bohr had made popular some decades before within his proposed model of the atom.} between the mathematically represented quantum superposition and the single observed measurement outcome \cite[p. 36]{Dirac74}. This new jump would become the famous ``collapse'' of the quantum wave function. During the posterior decades and up to our days, the need to justify Dirac's ``solution'' to the positivist dilemma became to be regarded as the most essential problem of QM itself. Something that the community of physicists and philosophers working in QM would name simply as ``the measurement problem''. 

It is important to notice that, from a mathematical viewpoint, instead of considering {\it matrices} right from the start ---as Heisenberg had done originally---, Dirac grounded the formalism of the theory only in terms of {\it vectors}. Let us recall that a unit vector, which in Dirac notation is written as a ket $|x\rangle$, can be seen as a rank one matrix through the following operation $|x\rangle\langle x|$. Indeed, if $\mathcal{H}$ is an $n$-dimensional complex vector space, $\mathcal{H}=\mathbb{C}^n$, and $B(\mathcal{H})$ is the space of $n\times n$ matrices, then we can relate the space of vectors $\mathcal{H}$ with the space of matrices $B(\mathcal{H})$ through the following map:
\[
\nu:\mathcal{H}\to B(\mathcal{H}),\quad \nu(|x\rangle):=|x\rangle\langle x|.
\]
Let us mention two relevant properties of $\nu$. The first one is that $\nu$ is not surjective. In fact,  its image is equal to the set of rank one matrices. The second relevant property of $\nu$ is that it is injective. Hence, we can think of the space of vectors as a subset of the space of matrices. In other words, the vector space is much ``smaller''\footnote{``Smaller'' in the sense that $\nu$ is an injection, and hence, the
space of vectors is included in the space of matrices.} than the matrix space. An important feature of the vectorial formulation is that, unlike the case of Heisenberg's matrix formulation, there always exist a basis in which the unit vector is represented by a single term ---instead of a superposition of them. And, of course,  following Born's rule, within this preferred basis the probability of observing this particular property is equal to unity. Consequently, unlike the general case of superpositions, such bases provide certain (probability = 1) knowledge of what will actually be the case if a measurement would be performed. 

Later on, this same idea\footnote{An idea which is also explicitly present in EPR's definition of their {\it element of physical reality}.} became explicitly introduced within the main postulates of the theory through the definition of the notion of {\it pure state}. As we have discussed in detail in \cite{deRondeMassri19b}, the notion of purity is operationally defined: `If a quantum system is prepared in such way that one can devise a maximal test yielding with certainty (probability = 1) a particular outcome, then it is said that the quantum system is in a \emph{pure state}.' In turn, the notion of {\it maximal test} allows to interpret a quantum observable as being an {\it actual property} ---i.e., a property that will yield the answer {\it yes} when being measured \cite{Smets05}. 
\begin{definition}[Operational Purity] Given a quantum system in the state $|\psi \rangle$,  there exists an experimental situation (or context) in which the test of it will yield with certainty (probability = 1) its related outcome. 
\end{definition}
However, as remarked in \cite{deRondeMassri19b}, there is also a mathematical definition according to which, the pure state of a quantum system is described by a unit vector, $|\psi \rangle$, in a Hilbert space. This definition has no physical counterpart and makes reference to a purely abstract mathematical feature of vectors, namely, that when considered in terms of density operators their norm is 1, that is, $\rho$ is a pure state if $\mbox{Tr}(\rho^2)=1$, or equivalently\footnote{A density matrix can be diagonalized, thus giving a set of eigenvalues $0\le\lambda_1\le\ldots<\lambda_n\le 1$ with $\sum_i \lambda_i = 1$. If $\mbox{Tr}(\rho^2)=1$, then $\lambda_1=\ldots=\lambda_{n-1}=0$ and $\lambda_n=1$. Hence, $\mbox{rk}(\rho)=1$ and then  $\rho=|v\rangle\langle v|$ and $\rho=\rho^2$. Conversely, if $\rho=\rho^2$ it has eigenvalues $0$ or $1$, but from $\sum_i \lambda_i = 1$ it follows $\lambda_1=\ldots=\lambda_{n-1}=0$ and $\lambda_n=1$. Hence, $\mbox{Tr}(\rho^2)=1$.} when $\rho=\rho^2$. 
\begin{definition}[Trace Purity]\label{pure} An abstract vector in Hilbert space $\Psi$ without any reference to a specific basis.\footnote{Like in \cite{deRondeMassri18b} we distinguish here between the purely abstract vector $\Psi$ and its specific representation in a basis $|\psi\rangle$.} Or in terms of density operators, an operator $\rho$ which is a projector where Tr$(\rho^2) = 1$ or $\rho = \rho^2$. 
\end{definition}
\noindent In this case the notion of pure state is obviously non-contextual, it is independent of the basis. The abstract vector $\Psi$ makes reference to the state $|\psi\rangle$, but also to any rotation $\sum a_i|\varphi\rangle_i$ (see for discussion \cite{daCostadeRonde16}). As remarked in \cite{deRondeMassri19b}, this latter definition is not consistent with the previous one. While the first operational definition ---which has the purpose to secure the existence of an observable which will be {\it certain} if measured--- is explicitly contextual (i.e. basis dependent) but not invariant, the  latter mathematically abstract definition of {\it pure state} is invariant (with respect to the trace) but lacks an operational content. 

\smallskip

Within the orthodox literature, in order to consider the original space of matrices containing all the operational content of the theory, the notion of pure state was extended to density operators. Let $\mathcal{H}$ be a Hilbert space. A density operator $\rho$  (i.e. a positive trace class  operator with trace 1) is called a \emph{state}. Being positive (and self-adjoint), the eigenvalues of $\rho$ are non-negative and real and it is possible to diagonalize it. If the rank of $\rho$ is equal to 1, this diagonal matrix is given by $(1,0,\ldots,0)$ and $\rho$ is equal to  $vv^{\dag}$ for some normalized vector $v\in\mathcal{H}$. In this case, $\rho$ is called a \emph{pure state}. For example, the vector  $\alpha|0\rangle+\beta|1\rangle$, $|\alpha|^2+|\beta|^2=1$, gives the following density matrix:  
\[
\rho=\begin{pmatrix}
|\alpha|^2 & \alpha\overline{\beta}\\
\overline{\alpha}\beta&|\beta|^2
\end{pmatrix}
\]
Notice that, if $\rho$ is a pure state (i.e., $\mbox{Tr}(\rho^2)=1$), there always exists a basis in which the matrix can be diagonalized as:
\[
\rho_{pure}=
\begin{pmatrix}
1 &0\\
0&0
\end{pmatrix}
\]

Given this link between vectors and matrices, it quite is easy to re-introduce the missing matrices of rank grater than 1 (or equivalently if $\mbox{Tr}(\rho^2)<1$). These were, of course, already present in Heisenberg's original formulation which made no distinction whatsoever between matrices and their rank. In the orthodox literature these re-introduced matrices became known as \emph{mixed states}; or in short, {\it mixtures}. Unlike the case of pure states, mixtures could not be represented as a unit vector, $|\psi \rangle$. Instead, mixed states were conceived as mixtures of pure states and mathematically represented as their convex sums:  
$$\rho_{mix} = \sum_{i} p_{i} \  \rho_{i}^{pure} =  \sum_{i} p_i \ | \Psi_i \rangle \langle\Psi_i | $$
Thus, while pure states guaranteed the existence of an observable which, if measured, would be obtained with {\it certainty} (probability equal to 1), mixed states did not. There is no single experimental context of measurement (i.e., no single basis) for which a mixed state will predict with certainty a {\it yes-no} answer for a specific observable. Mixtures provide uncertain knowledge regarding the pure state in which the quantum system is really supposed to be in. In this way, mixtures are able to introduce an ignorance interpretation of outcomes within the quantum formalism itself. As remarked by Nancy Cartwright \cite{Cartwright72}: ``The ignorance interpretation asserts that each member of the collection is in one of the pure states in the sum ---it is only our ignorance which prevents us from telling the right pure state for any specific member.'' Thus, contrary to the case of pure states, when considering mixed states, all observables are {\it uncertain}; they all possess a probability which pertains to the open interval $(0,1)$. But unlike quantum superpositions which possess {\it indeterminate} or {\it potential} properties, mixtures can be interpreted in terms of ignorance about pure states. As remarked by Cartwright: ``The ignorance interpretation is the orthodox interpretation for mixtures, and should not be confused with the ignorance interpretation for superpositions, which has been largely abandoned.'' In order to make things more explicit, we can consider as an example of a mixed state (i.e., $\mbox{Tr}(\rho^2)<1$) the following diagonal matrix,
\[
\rho_{mixed}=\begin{pmatrix}
\frac{1}{2} &0\\
0&\frac{1}{2}
\end{pmatrix}
\]

\noindent Here, the observables related to the diagonal elements have probability $\frac{1}{2}$ and consequently, it is orthodoxly claimed that this mixed state provides {\it minimal knowledge} about the actual quantum (pure) state in which the system really is. Before making the measurement, there is an equal 50 percent chance of obtaining either of the two possible results, so we do not know in which pure state the system is. 
 
As discussed in detail in \cite{deRondeMassri19b}, the just mentioned distinction between {\it pure states} and {\it mixtures} was introduced in order to support a twofold foundation. On the one hand, an empirical-positivist understanding of physics as an algorithmic mathematical device capable of predicting observations with certainty; and on the other, an atomist metaphysical understanding of physical reality in terms of systems constituted by definite valued properties. These two interconnected presuppositions have severe inconsistencies when related to the orthodox mathematical formalism of QM. It is only the contextual definition of {\it pure state} which provides the possibility to find out in a concrete case if the concept is {\it true} or {\it false}. Or in other words, it is only one basis between the infinitely many existent basis which contains a physical operational content. However, the mathematical non-contextual definition of {\it pure state}, which is not equivalent, lacks completely such operational reference. This is clearly problematic, for there is no obvious link between the contextual and the non-contextual definitions of {\it pure state}. Things become even more complicated when we shift our attention to {\it mixed states} in which case there is no clear operational counterpart beyond the reference to pure states --which are already ill defined.\footnote{While classical mixtures make reference to the ignorance of an underlying preexistent actual state of affairs, quantum mixtures ---due to the non-existence of a joint probability distribution \cite{Svozil17}--- are simply incompatible with such a (classical) ignorance interpretation; and even worse ---just like quantum superpositions \cite{deRonde18a}---, quantum mixtures lack a reference beyond measurement outcomes and mathematical structures.} The reference to `mixtures' ---as contra-posed to `pure states'--- has also become extremely problematic within the specialized literature, specially in the context of quantum information. It is a difficult problem to determine if a mixed state is separable or not \cite{LiQuiao18}; something which is essential within the research of quantum information processing given the orthodox definition of entanglement in terms of separability and purity. 

\smallskip

In the following sections we would like to turn our attention to the way in which two categorical approaches to QM have addressed ---in radically different ways--- the meaning of pure and mixed states. On the one hand, the topos approach and on the other, the more recent logos approach presented by the authors of this paper.

\section{Pure States and Mixtures in the Topos Approach}  

The topos approach, originally proposed by Chirs Isham, Jeremy Butterfield and Andreas D\"oring \cite{DoeringIsham08, DoeringIsham12, DoeringIsham11, Isham97, IshamButter1, IshamButter2, IshamButter3, IshamButter4}, presents a line of research that has been continued in different ways by researchers like Chris Heunen, Bas Spitters, Klas Landsman, Vasilios Karakostas and Elias Zafiris. The topos proposal can be regarded as a neo-Borhian categorical formulation which, taking as a standpoint the Kochen-Specker theorem, stresses the contextual character of QM. Motivated by Bohr's idea that the empirical content of quantum physics is accessible only through classical physics the topos approach attempts to provide a mathematical way to make sense of QM. But before entering the specific proposal of the topos, let us provide some basic mathematical notions regarding category theory.

\smallskip

First of all, a \emph{category} consists of a collection of \emph{objects} (often denoted as $X,Y,A,B$), a collection of \emph{morphisms} (or \emph{arrows}, denoted $f,g,p,q$) and four operations,
\begin{itemize}
\item To each arrow $f$, there exists an object $dom(f)$, called its \emph{domain}.
\item To each arrow $f$, there exists an object $codom(f)$, called its \emph{codomain}.
\item To each object $X$, there exists an arrow $1_X$, called the \emph{identity map} of $X$.
\item To each pair of arrows $f,g$ such that $dom(g)=codom(g)$ there exists a \emph{composition map}, $fg$ such that
$dom(fg)=dom(g)$ and $codom(fg)=codom(f)$.
\end{itemize}
An arrow $f$ is often denoted as $f:X\rightarrow Y$ to empathizes
the fact that $dom(f)=X$ and $codom(f)=Y$. 
We say that an arrow $f:X\rightarrow Y$ is invertible if there
exists an arrow $g:Y\rightarrow X$ such that $fg=1_Y$ and $gf=1_X$.
The collection of arrows between $X$ and $Y$ is denoted $\hom(X,Y)$. 
\begin{example}
The first example of a category is 
the category of sets $\mathcal{S}ets$.
Another example is the category of graphs.
The category of graphs, denoted $\mathcal{G}ph$,
extends naturally the category of sets.
A \emph{(simple) graph} is a set with a reflexive and symmetric relation. 
More formally, $\mathcal{G}$ is a graph if 
\begin{itemize}
\item Reflexivity: $P\sim P$ for all $P\in \mathcal{G}$.
\item  Symmetry: if $P\sim Q$, then $Q\sim P$ for all $P,Q\in \mathcal{G}$.
\end{itemize}
Elements of the graph are called \emph{nodes} and an \emph{edge} between
two nodes is present if these two nodes are related.
Arrows between graphs send nodes to nodes and edges to edges.
\end{example}
A remarkable fact is that the collection of categories has itself 
a structure of a category. The arrows are called \emph{functors}. 
A functor $F:\mathcal{C}\rightarrow \mathcal{D}$ assigns objects to
objects, arrows to arrows and is compatible with the
four operations (domain, codomain, identity and composition).

Let us present three standard constructions in category theory,
the \emph{comma category}, the \emph{graph of a functor}
and the \emph{category over an object}. The second construction
is a particular case of the first and the third of the second. Let $F:\mathcal{A}\rightarrow \mathcal{C}$ and  $G:\mathcal{B}\rightarrow \mathcal{C}$ be two functors with
the same codomain. 
The \emph{comma category} $F|G$
is a category whose objects are arrows in $\mathcal{C}$ of
the form 
\[
f:F(A)\rightarrow G(B),
\]
where $A\in\mathcal{A}$ and $B\in\mathcal{B}$. 
An arrow between $f$ and $g$ is a commutative square. The \emph{graph of a functor} $F:\mathcal{A}\rightarrow \mathcal{C}$ is defined as the comma category $F|1$,
where  $1=1_\mathcal{C}:\mathcal{C}\rightarrow\mathcal{C}$  is the identity functor. 
In the special case where
the functor $F$ is equal to $\hom(-,C)$ for some object 
$C\in\mathcal{C}$, the graph of this functor is called 
the category over $C$ and is denoted $\mathcal{C}|_C$.
This is our main structure. Objects in $\mathcal{C}|_C$
are given by arrows to $C$, $p:X\rightarrow C$, 
$q:Y\rightarrow C$, etc. Arrows $f:p\rightarrow q$
are commutative triangles,
\[
\xymatrix{
X\ar[rr]^f\ar[dr]_p& &Y\ar[dl]^q\\
&C
}
\]
\begin{example}
Let $\mathcal{S}ets|_\mathbf{2}$ be the category of sets
over $\mathbf{2}$, where $\mathbf{2}=\{0,1\}$ and $\mathcal{S}ets$ is
the category of sets.
Objects in $\mathcal{S}ets|_\mathbf{2}$
are functions from a set to $\{0,1\}$
and morphisms are commuting triangles, 
\[
\xymatrix{
\mathcal{G}_1\ar[rr]^f\ar[dr]_\Psi& &\mathcal{G}_2\ar[dl]^\Phi\\
&\{0,1\}
}
\]
In the previous triangle, $\Psi$ and $\Phi$ are objects of 
$\mathcal{S}ets|_\mathbf{2}$
and $f$ is a function satisfying $\Phi f=\Psi$.

As we mention before, this category is relevant in classical logic.
We can assign a true/false value to every element of $\mathcal{G}_1$
and/or $\mathcal{G}_2$ in a consistent manner. We say that $P\in \mathcal{G}_1$ 
(assume $P$ is a \emph{proposition} in the space $\mathcal{G}_1$) is true
if $\Psi(P)=1$, else we say that $P$ is false. 
Even more so, assume that we have a map $f:\mathcal{G}_1\rightarrow \mathcal{G}_2$
such that $\Phi f=\Psi$, then the truth or falsity of $P$ is unchanged
via $f$, that is $f(P)$ is true if and only if $P$ is true,
\[
f(P)\mbox{ is true }
\Longleftrightarrow 1=\Phi (f(P))=\Psi(P)
\Longleftrightarrow 
P\mbox{ is true }
\]
\end{example}
In the logos approach we generalize the category $\mathcal{S}ets|_\mathbf{2}$
extending $\mathcal{S}ets$ to $\mathcal{G}ph$ and the set $\mathbf{2}$
to the interval $[0,1]$.

\smallskip

Now that we have some basic facts and constructions from category theory, let
us review the topos approach and how it can 
handle mixed states.
For an introduction to the topos approach see \cite{deRondeMassri18a} or \cite{Eva}.
The topos approach makes use of the definition of \emph{context category} and of \emph{spectral presheaf}. Let $\mathcal{H}$ be a Hilbert space and consider $\mathcal{V}(\mathcal{\mathcal{H}})$ the set of commutative subalgebras $V\subseteq B(\mathcal{H})$ of bounded operators (with identity). Using the natural order in $\mathcal{V}(\mathcal{\mathcal{H}})$, we can consider it as a category. We call $\mathcal{V}(\mathcal{\mathcal{H}})$ the \emph{context category}. To each $V\in \mathcal{V}(\mathcal{\mathcal{H}})$  ($V\subseteq B(\mathcal{H})$ is  a commutative subalgebra with identity),  we assign its Gelfand spectrum (a compact topological space).  This assignment, denoted $\underline{\Sigma}$, is called the \emph{spectral presheaf}. The topos approach is particulary interested in some particular subobjects of $\underline{\Sigma}$. A subobject $\underline{S}$ is called \emph{clopen} if $\underline{S}(V)$ is a  clopen\footnote{A clopen subset in a topological space is a set both open and closed.} subset of $\underline{\Sigma}(V)$ for all $V\in\mathcal{V}(\mathcal{\mathcal{H}})$. We denote the set of clopen-subobjects as $Sub_{cl}(\underline{\Sigma})$. The authors construct a map $\delta:\mathcal{P}(\mathcal{H})\to Sub_{cl}(\underline{\Sigma})$ called \emph{daseinisation of projection operators} which sends each projector $P_i$ to a clopen subobject $\delta(P)$, where $\mathcal{P}(\mathcal{H})$ denotes the set of projectors in $\mathcal{H}$. 
The basic idea behind this construction is to recover classical physics. 
For each $V\in \mathcal{V}(\mathcal{\mathcal{H}})$, the space 
$\underline{\Sigma}(V)$ has to be interpreted as a state space and for each
projector $P$, the subset $\delta(P)(V)$ has to be interpreted as a 
proposition in the state space $\underline{\Sigma}(V)$. In summary, according to the slogan 
\emph{quantum physics is equivalent to classical physics
in the appropriate topos}, \cite{DoeringIsham12}, 
the topos approach defines for each 
$V\in \mathcal{V}(\mathcal{\mathcal{H}})$, 
a state space and a Boolean logic (all subject to compatibilities conditions).

In \cite{Doering11, DoeringIsham12} Isham and Do\"ering make an attempt to incorporate to the topos approach the notions of probability and density matrices. In order to do so, they extend their previous constructions \cite[p. 6]{DoeringIsham12}. As they \cite[p. 3]{DoeringIsham12} argue: ``Probabilities are thereby built into the mathematical structures in an intrinsic manner. They are tied up with the internal logic of the topos and do not show up as external entities to be introduced when speaking about experiments'' The authors define the presheaf  $\underline{[0,1]}^{\ge}$ given by $[0,1]$-valued, nowhere-increasing functions on $\mathcal{V}(\mathcal{\mathcal{H}})$. In other words, if $p\in \underline{[0,1]}^{\ge}$, then $p(V)\in[0,1]$ for all $V\in\mathcal{V}(\mathcal{\mathcal{H}})$ and if $V'\subseteq V$ we have $p(V')\ge p(V)$. Now, given a density matrix $\rho$, the authors constructed a map $\mu^{\rho}:Sub_{cl}(\underline{\Sigma})\to\underline{[0,1]}^{\ge}$ such that, when  restricted to the image of $\delta$ and taking minimum over $\mathcal{V}(\mathcal{\mathcal{H}})$, they recover Born's rule,
\[
\min_{V\in \mathcal{V}(\mathcal{\mathcal{H}})}
\mu^{\rho}(\delta(P))(V):=\mbox{Tr}(\rho P)
\]
The general definition of $\mu^{\rho}$ to the whole set
$Sub_{cl}(\underline{\Sigma})$ is rather technical and non-trivial.
In fact, several alternative constructions are needed in order to prove
the previous formula. The basic idea behind their construction
is that a density matrix $\rho$ defines a probability measure
on each state space $\underline{\Sigma}(V)$.

At this point it becomes important to make some remarks about the topos program. Even though the authors of the present paper believe the topos approach is a very interesting and original proposal, it has several mathematical and philosophical drawbacks. From a mathematical point of view, it is evident that it is necessary to adapt all the previous formalism to the new mathematical constructions in order to be able to incorporate mixed states. The result of this process is the creation of a very complex and elaborated theory which is difficult to follow even for an expert in the mathematical field. Furthermore, it is not even clear if the results in the previous formulation are still valid. But also from a philosophical point of view there seems to exist a tension between, on the one hand, a supposedly realist approach to physics which attempts to talk about systems with well defined properties,\footnote{As remarked by D\"oring and Isham in \cite{DoeringIsham08}: ``When dealing with a closed system, what is needed is a realist interpretation of the theory, not one that is instrumentalist. The exact meaning of `realist' is infinitely debatable, but, when used by physicists, it typically means the following: (1) {\it The idea of  ``a property of the system" (i.e. ``the value of a physical quantity'') is meaningful, and representable in the theory.} (2) {\it Propositions about the system are handled using Boolean logic. This requirement is compelling in so far as we humans think in a Boolean way.} (3) {\it There is a space of ``microstates'' such that specifying a microstate leads to unequivocal truth values for all propositions about the system. The existence of such a state space is a natural way of ensuring that the  first two requirements are satisfied.} The standard interpretation of classical physics satisfies these requirements, and provides the paradigmatic example of a realist philosophy in science. On the other hand, the existence of such an interpretation in quantum theory is foiled by the famous Kochen-Specker theorem."} and on the other, a neo-Bohrian scheme\footnote{The topos approach has been developed explicitly as a neo-Bohrian attempt to understand QM. A reference to the Danish physicist which has become completely explicit not only in the works of Chris Heunen, Klaas Landsman and Bas Spitters, but also in the works of Vasilios Karakostas and Elias Zafiris.} which makes explicit use of several anti-realist ideas (e.g., that `reality is contextual').

\section{Beyond Purity and Mixtures in the Logos Approach}

While the topos approach is considered to be a neo-Bohrian approach, by following some of the main ideas present within the works of Einstein, Heisenberg and Pauli ---specifically in relation to their understanding of physical theories--- the logos presents a line of research exactly in the opposite direction. Indeed, within the logos, taking as a standpoint the orthodox mathematical formalism we attempt to restore an objective-invariant account of QM in which there are no {\it preferred bases} or {\it factorizations} and subjects become ---as Einstein required--- completely {\it detached} from the theoretically represented state of affairs. We do so by willingly paying the price of abandoning the classical metaphysical account of reality in terms of `systems' and `properties' ---which the topos approach, following Bohr, wants to retain at al costs--- and introducing a new non-classical representation in which the physical notions of {\it power} and {\it potentia} play an essential role. By staying close to the operational invariance of the Born rule, our main interest becomes the category $\mathcal{G}ph|_{[0,1]}$ of graphs over the interval $[0,1]$. Let us begin by reviewing some properties of the category of graphs. First, we give an example of a graph coming from the quantum formalism,
\begin{example}
Let $\mathcal{H}$ be Hilbert space and let $\Psi$ be a vector, $\|\Psi\|=1$.
Take $\mathcal{G}$ as the set of observables with the commutation relation
given by QM, the \emph{quantum commutation relation}.
This relation is reflexive, symmetric but \textbf{not} transitive, hence
$\mathcal{G}$ is a non-complete\footnote{A graph is complete if there is an edge between two arbitrary nodes.} graph. 
\end{example}
\begin{definition}
Let $\mathcal{G}$ be a graph.
A \emph{context} is a complete subgraph (or aggregate) inside $\mathcal{G}$. A \emph{maximal context} is a context not contained properly in another context. If we do not indicate the opposite, when we refer to contexts we will be implying maximal contexts.
\end{definition}
\noindent For example, let $P_1,P_2$ be two nodes of a graph $\mathcal{G}$. 
Then, $\{P_1, P_2\}$ is a context if $P_1$ is related to $P_2$, $P_1\sim P_2$. Saying differently, if there exists an edge between $P_1$ and $P_2$. In general, a collection of nodes $\{P_i\}_{i\in I}\subseteq \mathcal{G}$ determine a {\it context} if $P_i\sim P_j$ for all $i,j\in I$. Equivalently, if the subgraph with nodes $\{P_i\}_{i\in I}$ is complete. 
\begin{theo}
Let $\mathcal{H}$ be a Hilbert space and let $\mathcal{G}$ be the graph of immanent powers with the commutation relation given by QM. It then follows that: 
\begin{enumerate}
\item The graph $\mathcal{G}$ contains all the contexts (or quantum perspectives). 
\item Each context is capable of generating the whole graph $\mathcal{G}$.
\end{enumerate}
\end{theo}
\begin{proof}
See \cite{deRondeMassri18b}.
\qed
\end{proof}

\smallskip 
\smallskip 

In the logos approach we work with the category
$\mathcal{G}ph|_{[0,1]}$.
An object in $\mathcal{G}ph|_{[0,1]}$ consists of a 
map $\Psi:\mathcal{G}\rightarrow [0,1]$, where $\mathcal{G}$ is a graph. 
Intuitively, $\Psi$
assigns a \emph{potentia} to each node of the graph $\mathcal{G}$.
Specifically, to each node 
$P\in\mathcal{G}$, we assign a number $\Psi(P)$, but this time, 
$\Psi(P)$ is a number between $0$ and $1$.
Then, in order to provide a map to the graph of immanent powers, we use the Born rule. We remark that in the logos scheme the Born rule is not an axiom added to the theory which would require an independent derivation  ---as argued, for example, by Deutsch and Wallace \cite{Deutsch99, Wallace07}--- but a consequence of the orthodox mathematical formalism itself. Gleason's theorem \cite{Gleason57} is just an answer to the mathematical problem of defining all measures on the closed subspaces of a Hilbert space. Gleason's theorem derives the Born rule as the natural measure for QM, and at the same time precludes the possibility of two valued measures ---which is also related to the famous result by Kochen and Specker (see \cite{Svozil17, deRondeMassri18a} for a detailed analysis). Thus, to each \emph{power} $P\in\mathcal{G}$, we assign through the Born rule  the number $p=\Psi(P)$, where $p$ is a number between $0$ and $1$ called \emph{potentia}. As discussed in detail in \cite{deRondeMassri18a}, we call this  map $\Psi:\mathcal{G}\rightarrow [0,1]$ a {\it Potential State of Affairs} (PSA for short). Summarizing, we have the following:
\begin{definition}
Let $\mathcal{H}$ be a Hilbert space and let $\rho$ be a density matrix.
Take $\mathcal{G}$ as the graph of immanent powers with the quantum commutation relation. 
To each immanent power $P\in\mathcal{G}$ apply the Born rule to get the number $\Psi(P)\in[0,1]$, which is called the potentia (or intensity) of the power $P$.  Then, $\Psi:\mathcal{G}\rightarrow [0,1]$
defines an object in $\mathcal{G}ph|_{[0,1]}$. We call this map a \emph{Potential State of Affairs} (or a \emph{PSA} for short).
\end{definition}
\noindent Intuitively, we can picture a PSA as a table,
\[
\Psi:\mathcal{G}(\mathcal{H})\rightarrow[0,1],\quad
\Psi:
\left\{
\begin{array}{rcl}
P_1 &\rightarrow &p_1\\
P_2 &\rightarrow &p_2\\
P_3 &\rightarrow &p_3\\
  &\vdots&
\end{array}
\right.
\]

\noindent Thus, an abstract vector in Hilbert space (or equivalently, a density matrix) without refeernce to a basis provides a table of intensive powers describing an objective PSA. In this frame, the Born rule, contrary to the orthodox interpretation ---followed also by the topos approach \cite{DoeringIsham12}---, acquires an objective reference, namely, the intensive invariant measure (i.e., the potentia) of all quantum powers. As stressed in the logos approach, the experimental account of such intensive quantification must be obviously acquired through a statistical analysis. Unlike in the orthodox interpretation of the Born rule, a single measurement outcome cannot be regarded as the meaningful reference of theoretical prediction. Single outcomes simply do not provide enough information to consistently refer to the number $p$ which pertains to the interval $(0,1)$. Or in other words, a single measurement result cannot provide the value of the potentia of a power, we always must require many measurements of the same power in order to estimate its intensity (see \cite{deRonde16a, deRondeFreytesSergioli19}).  

In this way, the logos approach departs from the well known positivist idea according to which physical theories predict measurements that can be restricted to {\it yes-no} elementary tests. This idea is explained by Asher Peres \cite[p. 202]{Peres02} in the following way: ``We start with some definitions and propositions which are not controversial. {\it There are `elementary tests' (yes-no experiments) labelled A, B, C, . . . Their outcomes are labelled a, b, c, ... = 1 (yes) or 0 (no).} In quantum theory, these elementary tests are represented by projection operators.'' It is this idea ---considered by many as ``not controversial''--- which implies the (metaphysical) imposition of a {\it binary} reference to physical existence. After this definition, there is also a ---very controversial--- shift from the empirical finding of actual measurement outcomes (observables) to the metaphysical reference of projection operators now understood as preexistent properties of quantum systems (see section 2 and also \cite{deRonde18}). As Peres continues to explain:
\begin{quotation}
\noindent {\small ``The simplest observables are those for which all the coefficients $a_r$ are either 0 or 1. These observables correspond to tests which ask yes-no questions (yes = 1, no = 0). They are called {\it projection operators}, or simply projectors, for the following reason: For any normalized vector $v$, one can define a matrix $P_v = vv^{\dag}$, with the properties $P_v^2 = P_v$ and $P_v u = vv^{\dag} u = v \langle v,u \rangle$ (3.52) 
\noindent The last expression is a vector parallel to $v$, for any $u$, unless $\langle v,u \rangle = 0$. In geometric terms, $P_v u$ is the projection of $u$ along the direction of $v$.'' \cite[p. 66]{Peres02}}
\end{quotation}

\noindent It is by imposing this binary restriction to the values of projection operators that, as explicitly shown by the Kochen-Specker theorem \cite{KS}, one reaches a contradiction. Like many others today, Peres \cite[p. 14]{Peres02} concludes that therefore: ``Quantum physics [...] is incompatible with the proposition that measurements discover some unknown but preexisting reality.'' This conclusion goes back to Bohr's analysis of QM and his insistence that the most important (epistemological) lesson to be learnt from QM is that, we subjects, are not only spectators but also actors in the great drama of (quantum) existence. However, as we have demonstrated in \cite{deRondeMassri18a} through the explicit development of an intensive non-contextuality theorem, this is simply not true. When considering the Born rule as computing {\it intensive values} ---something which is natural from a mathematical physical perspective which understands that invariance is the key for any consistent representation---, objectivity can be restored and QM becomes compatible with the proposition that (statistical) measurements discover an unknown but preexistent (potential) reality. Of course, the price we have willingly paid is to give up the classical (metaphysical) representation of reality in terms of actual `systems' and `properties'. An idea which regardless of its serious ---both mathematical and conceptual--- difficulties, has been dogmatically retained in almost all interpretations of QM.  

\smallskip

To sum up, some important remarks go in order: 
\begin{enumerate}
\item[I.] Our Logos approach makes explicit the existence of two distinct levels of mathematical representation regarding vectors. On the one hand, we have the PSA, i.e., an abstract vector in Hilbert space $\Psi$; and on the other hand, we have the particular basis-representation of the PSA in a specific context; i.e., the vector written in a basis $|\psi \rangle$ which we call a quantum perspective. While the first level is obviously non-contextual, the second level is explicitly contextual or basis dependent (see for a detailed analysis \cite{deRondeMassri18b}). In the logos approach we have not only different names for these different concepts but also a notation which makes explicit this fundamental distinction right from the start. While we use capital Greek letters, e.g. $\Psi$, to refer to an abstract vector (or PSA), we apply Dirac's notation, e.g. $|\psi \rangle$, when making reference to a specific experimental context. 

\item[II.] The interpretation of the Born rule as an intensive quantification of powers (projection operators) allows to bypass the need of a {\it binary valuation}. But more importantly, it provides a {\it global intensive valuation} which escapes the constraints of the Kochen-Specker theorem allowing to restore an objective reference to the quantum formalism \cite{deRondeMassri18a}.  

\item[III.] The logos approach embraces the shift from a {\it binary} understanding of {\it certainty} to an {\it intensive} one. From this standpoint, the number that we find by applying the Born rule is not a measure of `lack of knowledge' of an inaccurate representation of an {\it Actual State of Affairs} ---which QM denied can be given in the first place. It is, on the very contrary, an objective account of the potentia of the powers constituting an objective and invariant {\it Potential State of Affairs}. As a direct consequence, the distinction between {\it pure state} and {\it mixed state} becomes completely irrelevant \cite{deRonde16a, deRondeFreytesSergioli19}.  
\end{enumerate}

\section{The Invariant-Operational Role of Bases in QM} 

As it is well known, Bohr's contextual interpretation of QM was presented within his influential reply \cite{Bohr35} to the famous EPR paper \cite{EPR}. Bohr imposed as a pre-requisit, in order to make reference to the properties of a quantum system, the need to specify the context of measurement ---something that was codified, in mathematical terms, in the choice of a (preferred) basis or complete set of commuting observables. He presented this requirement knowing already that it was not possible to consistently assign a {\it global (binary) valuation} to all properties of a quantum system. As it was early remarked by Schr\"odinger \cite[p. 156]{Schr35}: ``[...] if I wish to ascribe to the model at each moment a definite (merely not exactly known to me) state, or (which is the same) to {\small {\it all}} determining parts definite (merely not exactly known to me) numerical values, then there is no supposition as to these numerical values {\small{\it to be imagined}} that would not conflict with some portion of quantum theoretical assertions.'' Bohr had already welcomed the idea that: ``We must, in general, be prepared to accept the fact that a complete elucidation of one and the same [quantum] object may require diverse points of view which defy a unique description.'' Consequently, his interpretational maneuvers were not intending to find a consistent representation for the theory of quanta. That was a task he considered hopeless. The lack of unity and consistency within theoretical representation implied a silent but radical shift of the role ---and understanding--- of reference frames (or bases) in physical theories. Following Bohr's contextuality, the choice of a {\it preferred basis} begun to be regarded as conditioning the possibility to refer to a specific sub-set of properties of a quantum system. This idea was advanced in more rigorous mathematical terms by Dirac's famous presentation of the theory in 1930 \cite{Dirac74}. Let us discuss this kernel point in some detail.

At least before Bohr's contextual interpretation of QM, mathematical invariance played an essential role in all physical theories providing the conditions of possibility for a subject-independent representation of a state of affairs. Physical theories provided on the one hand, an abstract definition of objects, and on the other, a way to conceive the particular {\it states} of these objects (or systems) through the specification of a {\it reference frame} (or basis) which, in turn, could be related to a particular observation. In theoretical physics, while the abstract definition of an object is basis independent, its connection to experience always requires the specification of a reference frame in which the specific values of the properties characterizing the object can be analyzed accordingly. Thus, one thing is to provide the definition, in general terms, of a physical object or {\it system}, say a `field' or a `particle', and a very different thing is to consider the specific {\it state} of a `field' or a `particle' within a given situation, something which necessarily requires the consideration of a {\it reference frame}. It is only from the choice of a specific reference frame that the more specific representation of the state can be provided. In physics while there can be an abstract representation of an object, there can be no representation of its state without a reference frame. For example, one might think of a dog as an object in space-time with properties such as position and momentum. Let us consider the situation of a dog running across a street. As it is well known, the representation of this situation will obviously depend on the particular frame of reference. While the velocity of the dog might be $v_d$ for someone next to it drinking a coffee in a bar, it will differ from the velocity described by someone in a car $v_d - v_c$. What is essential is that in order to give {\it consistency} to the many possible basis dependent representations of the dynamical properties of the dog, classical mechanics makes use of the so called Galilean transformations. These transformations secure an invariant type representation of the state of the system in such a way that all reference-frame dependent representations can be regarded as {\it different} and yet referring to the {\it the same} state of the object. The important aspect introduced by this type of representational-invariance is that it allows, through the transformations, to provide a consistent translation between the multiple basis-dependent theoretical representations of {\it the same} state of {\it the same} object. In physical theories all this is also formulated in mathematical terms. A system considered in terms of a set of actual (definite valued) properties can be thought as a map from the set of properties to the $\{0,1\}$. More specifically, a system  $\mathcal{S}$ might be regarded as a function $\Psi: \mathcal{S}\rightarrow\{0,1\}$ from the set of properties to $\{0,1\}$ satisfying certain compatibility conditions (see \cite{deRondeMassri18a}).  We say that the property $P\in \mathcal{S}$ is \emph{true} if $\Psi(P)=1$ and  $P\in \mathcal{S}$ is \emph{false} if $\Psi(P)=0$. The description of a system with respect to a reference frame is formalized by the fact that the morphism $G$ satisfies $\Phi G=\Psi$. Diagrammatically, 
\[
\xymatrix{
\mathcal{S}_{B_1}\ar[dr]_{\Psi}\ar[rr]^G&&\mathcal{S}_{B_2}\ar[dl]^{\Phi}\\
&\{0,1\}}\]
The transformation does not only provide the properties which are true but also the consistent values of the {\it dynamical properties} such as position, velocity, etc.,
\[
\xymatrix{
P^i{_{B_1}}\ar@{|->}[d]&
\mathcal{S}_{B_1}\ar[dr]\ar[rr]^G&&\mathcal{S}_{B_2}\ar[dl]&P^i{_{B_2}}\ar@{|->}[d]\\
v(P^i{_{B_1}})&&\mathbb{R}   &&v(P^i{_{B_2}})
} 
\]
Let us explain these diagrams in more detail. The set $\mathcal{S}_{B_1}$ consists of the dynamical properties 
of the system $\mathcal{S}$ in the reference frame $B_1$. The map from $\mathcal{S}_{B_1}$ to $\mathbb{R}$ sends
a property $P^i_{B_1}$ to its real value $v(P^i{_{B_1}})$
(for example, the velocity or the position). Same situation with $\mathcal{S}_{B_2}$.
The fact that the diagram in the middle commutes implies that the values of the properties in the reference frame $B_1$ are mapped
under $G$ in a compatible way to values in the reference frame $B_2$. For example, the velocity of a car viewed from a speed train will have a different but compatible value, under the transformation $G$, with respect to the velocity of the same car viewed from the train-station. Thus, given the representation of the state of the system $\mathcal{S}_{B_1}$ in a particular reference frame, $B_1$, we can imply through the transformation $G$, how the (same) state of the system will be consistently represented from any other frame of reference, $B_2$, as $\mathcal{S}_{B_2}$. This type of invariance can be also extended for the dynamical properties of the system (see \cite{deRondeMassri16}). As discussed in \cite[Sect. 1]{deRondeMassri18a}, the conjunction of systems, $\mathcal{S}_i$ (with $i =1...n$), can also give rise to a consistent {\it Actual State of Affairs} (ASA) composed of many systems. Due to the value-invariance of the properties with respect to reference frames, the context of experimental analysis (reference frame or basis) becomes completely irrelevant for the theoretical characterization of the real physical situation. It is this particular invariant feature of physical representation which allows to talk about a real state of affairs, independent of any subjective perspective or viewpoint. And as Einstein would constantly remark, it is also this which allows to regard any particular empirical subject (linked to a mathematical basis) as {\it detached} from the representation of the whole course of events.  

Of course, this does not mean that reference frames are unimportant in physics. On the very contrary, in theoretical physics the reference frame is essential in order to describe any situation or experiment in a lab. Actual experiments are not just abstract mathematical or conceptual representations, they require the specification of the conditions under which the testing takes place and becomes possible. The most important of these conditions is the reference frame itself. The position of an object has meaning only with respect (or relative) to a reference frame. Without the specification of a viewpoint it is impossible to refer to the state of a system. And this is the reason why, the {\it Meaningful Operational Statements} produced by theories are always basis-dependent. They cannot be given without the specification of a reference frame (or basis). 

\smallskip 
\smallskip 

\noindent {\bf Meaningful Operational Statement:} {\it Every operational statement within a theory capable of predicting the outcomes of possible measurements must be considered as meaningful with respect to the representation of physical reality provided by that theory in connection to a specific frame of reference or basis.}

\smallskip 
\smallskip 

\noindent Any theory which attempts to provide an objective (subject-independent) representation of a state of affairs must be able to refer to such meaningful operational statements in an invariant manner, that is, in a way in which all possible results of experiments can be consistently pictured as a consequence of a preexistent theoretically represented {\it moment of unity} ---e.g., the state of an object. The {\it many} reference frames dependent representation must be able to provide a consistent account of {\it the same} state of affairs. This is the essential content of the Greek meaning of understanding ---which they often related to the relation between the one and the many. As Heisenberg \cite[p. 63]{Heis71} explained: ```Understanding' probably means nothing more than having whatever ideas and concepts are needed to recognize that a great many different phenomena are part of coherent whole.'' On the contrary, leaving behind {\it objectivity}, as the search for a consistent unity in the representation of phenomena, Bohr embraced an {\it intersubjective}\footnote{Bohr systematically referred to intersubjective statements as objective ones.} reference to observable phenomena and stressed that it was the communication between subjects that which allowed to avoid any sort of ambiguity \cite[p. 98]{D'Espagnat06}: ``The description of atomic phenomena has in these respects a perfectly objective character, in the sense that no explicit reference is made to any individual observer and that therefore... no ambiguity is involved in the communication of observation." Bohr shifted the meaning of `experiment' from the consistent representation of a real state of affairs to the possibility of communication between subjects. As he \cite[p. 7]{WZ} argued ``the very word {\small {\it experiment}} refers to a situation where we can tell others what we have done and what we have learned." In this way, placing measurement as a standpoint, Bohr explicitly rejected the very possibility of providing a consistent picture of a state of affairs independently of the choice of the basis (or reference frame). As it was later on explicitly shown through KS theorem \cite{deRondeMassri16, KS}, he was right to point out that if attempting to restore a binary representation in terms of definite valued properties of systems the enterprise would be doomed to failure, but he was wrong to argue that this could not be avoided in general. As it was explicitly shown by the logos approach \cite{deRondeMassri18a} the consideration of an {\it intensive valuation} is able to advance a global consistent invariant account of the mathematical formalism which bypassing the KS theorem, in turn, also allows to provide an objective (subject-independent) representation of the state of affaires described by the theory of quanta.

\smallskip

As mentioned above, and explicitly analyzed in \cite{deRondeMassri19b}, since Dirac's textbook formulation there has been a deep problem in order to address the meaning of {\it states} and {\it systems} in QM. Even today, there still exists within the literature a deep and widespread disorientation related to the meaning of these notions which are commonly applied in inconsistent manners sometimes referring to abstract mathematics while others to ``small particles'' (see \cite[Chap. 4]{Schlosshauer11}). We believe that the reason behind this situation is the inadequate link, introduced by Dirac, between the mathematical formalism and these physical notions. While in physics a {\it state} makes reference to the specific features of an object, Dirac presented an interpretation of the mathematical formalism in which each {\it state} was related to an observation and represented by a {\it ket} vector. As a consequence, it made perfect sense to argue that {\it the same system} could be represented in terms of {\it different states}, depending on the basis: 
\begin{quotation}
\noindent {\small ``[E]ach state of a dynamical system at a particular time corresponds to a ket vector, the correspondence being such that if a state results from the superposition of certain other states, its corresponding ket vector is expressible linearity in terms of the corresponding ket vectors of the other states, and conversely. Thus the state $R$ results from a superposition of the states $A$ and $B$ when the corresponding ket vectors are connected by $ | R \rangle\  = c_1 | A \rangle\ + c_2 | B \rangle\ $.'' \cite[p. 16]{Dirac74}}
\end{quotation}
In this way, the meaning of {\it sameness} as related to reference frames became completely inverted. Reference frames (or bases) did not make anymore reference to {\it the same} state of {\it the same} object, but instead to {\it different states} ---and consequently, different observables. The presupposition that a system could only have a single state at a time was then left behind as a classical prejudice. Each basis made then reference to a specific set of possible states of a system. The state, or superposition of states, was determined by the choice of the basis regardless of any invariant theoretical condition between them. Dirac ---in a Bohrian fashion--- simply blamed the situation on quantum superpositions for being non-classical: ``The nature of the relationships which the superposition principle requires to exist between the states of any system is of a kind that cannot be explained in terms of familiar physical concepts. One cannot in the classical sense picture a system being partly in each of two states and see the equivalence of this to the system being completely in some other state.'' According to Dirac, following Bohr, this implied a limit in the possibility of representation. But this was not a problem. Dirac had already ``remarked that the main object of physical science is not the provision of pictures, but the formulation of laws governing phenomena and the application of these laws to the discovery of phenomena. If a picture exists, so much the better; but whether a picture exists of not is a matter of only secondary importance.'' As he also made the point, it is ``important to remember that science is concerned only with observable things'', consequently, the reference to superpositions should be regarded only as an abstract mathematical game which does not need to refer to anything real ---even though QM talks about small particles. What Dirac did not seem to recognize ---maybe because he was not a physicist--- was that his interpretation implied a much deeper commitment, one which precluded the possibility for QM of an invariant operational discourse. The abstract invariance of unit vectors with respect to the trace seemed enough to Dirac. It is interesting to notice that during the 1940s and 1950s Dirac's state-vectorial formulation was established in textbook QM through the essential introduction of the notion of {\it pure state}. A notion which makes even more explicit the double reference, on the one hand, to the abstract invariance of a unit vector, and on the other, to the non-invariant certain prediction of a single observable (see for a detailed discussion \cite{deRondeMassri19b}).\footnote{Today, the reference to states has become explicitly instrumental. As pointed out by Chris Timpson, for many contemporary researchers, the quantum state does not represent ``how things are in an external, objective world, it merely represents what information one has. Mermin (2001), Peierls (1991), Wheeler (1990) and Zeilinger (1999) have all endorsed this kind of view. Hartle (1968, p. 709) provides an excellent summary: {\it `The state is not an objective property of an individual system but is that information, obtained from a knowledge of how a system was prepared, which can be used for making predictions about future measurements'.}''}   

\smallskip

Escaping from these orthodox set of definitions and going back to an analysis of the mathematical formalism as related to its invariant-operational ---and representational--- conditions, an outmost important feature of our logos approach is that it allows us to distinguish explicitly between these two just mentioned levels of mathematical representation through the notions of {\it Potential State of Affairs} (PSA) and {\it Quantum Perspective}. While the PSA provides a purely abstract basis independent representation, the quantum perspective makes explicit reference to a specific context or basis dependent representation of that {\it same} PSA. Due to the operational-invariance of the Born rule, all quantum perspectives can be considered simultaneously without the inconsistencies found in Dirac's formulation ---which extends the number of states not only for different bases, but also, even for the same representation. Consequently, we are able to restore an objective representation in which measurements can be regarded as epistemic forms of testing the physical representation of a (potential) state of affairs. In order to make explicit these different levels of physical representation the logos approach introduces a new important distinction within Dirac's notation. While abstract vectors ---which refer to a PSA--- are noted with capital Greek letters, we retain Dirac's notation in order to refer to vectors written in a specific basis ---which we interpret as a quantum perspective. Given a PSA, $\Psi$, defined by an abstract unit vector, $v$, and given a basis (or context), $\mathcal{C}=\{|w_1\rangle,\ldots,|w_k\rangle\}$, we can write $v$ as a quantum superposition or, as we shall now call it for reasons that will become evident, a {\it Quantum Perspective}: 
\[ QP_{\Psi,\mathcal{C}}:=\sum_{i=1}^k c_i|w_i\rangle. \]
In fact, we can assign to $\Psi$ a multiplicity of different quantum perspectives (or superpositions):
\[
QP_{\Psi,\mathcal{C}_1},QP_{\Psi,\mathcal{C}_2},\ldots,QP_{\Psi,\mathcal{C}_n}
\]
one for each context $\{\mathcal{C}_1,\ldots,\mathcal{C}_n\}$. Even more, as remarked in Theorem 5.2, each quantum superposition can {\it generate} ($\rightsquigarrow$) not only the other superpositions (by simply making a change of basis) but also the whole PSA,
\[
QP_{\Psi,\mathcal{C}_1}\rightsquigarrow
QP_{\Psi,\mathcal{C}_2}\rightsquigarrow 
\ldots\rightsquigarrow
QP_{\Psi,\mathcal{C}_n}\rightsquigarrow
\Psi.
\]
It is also true, as already remarked in \cite{daCostadeRonde16}, that from a mathematical viewpoint there is a class of equivalence between the different representations which allow us to write the following equalities: 
\[
QP_{\Psi,\mathcal{C}_1}=
QP_{\Psi,\mathcal{C}_2}= 
\ldots=
QP_{\Psi,\mathcal{C}_n}=
\Psi.
\]
This {\it equivalence relation} obviously does not imply that these different quantum perspectives are making reference to a physical system constituted by properties. In fact, there is no {\it Global Binary Valuation} of the properties considered from different reference frames (for a more detailed discussion see \cite{daCostadeRonde16, deRondeMassri16, deRondeMassri18a, deRondeMassri18b}).

A useful visual account of the just mentioned definitions and levels is provided in the logos approach through the use of {\it graphs} in which the partially filled nodes represent {\it powers} with their respective {\it potentia}. Graphs allow us to picture simultaneously the whole PSA, $\Psi$, the different context dependent quantum perspectives, $QP_{\Psi, {\mathcal C_i}}$, as well as each different power with its respective invariant potentia computed via the Born rule. Let us stress the fact that in this new account of the formalism each node ---which has a one to one relation to a {\it ket}--- singles out a {\it power}, something that in Dirac's formulation was related to a {\it state} or an {\it observable}. While in the logos approach, given an abstract unit-vector, all kets (i.e., all powers) make reference to the same (potential) state of affairs (quantum perspectives being just particular basis-dependent representations), in Dirac's formulation each ket singles out a particular state and each different basis a set of them, all of them different between each other. While in Dirac's presentation there is no global consistency, in the logos approach all quantum perspectives can be consistently referred ---through the operational-invariance of the Born rule--- to the same {\it Potential State of Affairs} ---restoring in this way an objective (subject-independent) representation.     
\begin{center}
\includegraphics[width=19em]{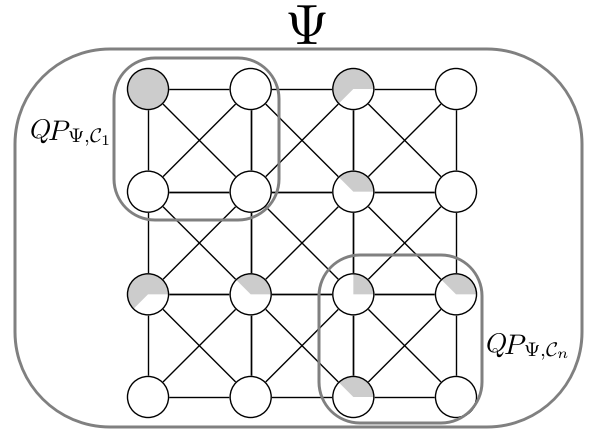}
{\small \fig: A PSA, $\Psi$, with two different Quantum Perspectives, {\it 1} and {\it n}, pointed out.}
\end{center}
In figure 1 we can clearly see that even though there is a sense in which $QP_{\Psi,\mathcal{C}_1}= QP_{\Psi,\mathcal{C}_n}$, there is also an obvious sense in which $QP_{\Psi,\mathcal{C}_1} \neq QP_{\Psi,\mathcal{C}_n}$. An obvious difference between them is that the quantum perspectives $QP_{\Psi,\mathcal{C}_1}$  and $QP_{\Psi,\mathcal{C}_n}$ are not making reference to the same section of the graph $\Psi$. According to the logos approach, each one of these different quantum perspectives is making reference to a different subset of powers and consequently to different experimental arrangements. Each one of them provides a set of {\it meaningful operational statements} related to the statistical testing of powers and their specific potentia. Each quantum perspective provides thus, objective (intensive) information of the same (potential) state of affairs. 

Without loosing any generality, it is interesting to notice that the following theorem guarantees that the PSA representation is equivalent to the density matrix representation:
\begin{theo}
The knowledge of a particular PSA, $\Psi$, is equivalent to the knowledge of the density matrix $\rho_{\Psi}$. In particular, if $\Psi$ is defined by a normalized vector $v_{\Psi}$, $\|v_{\Psi}\|=1$, then we can recover the vector from $\Psi$.
\end{theo}
\begin{proof}
See \cite{deRondeMassri18b}.
\qed
\end{proof}

\smallskip
\smallskip

\noindent As remarked above, reference frames are essential in making explicit the operational content of a theory. Through the use of graphs the logos representation is capable to account for both non-contextual and contextual levels simultaneously. While the whole graph provides an account of the non-contextual PSA, $\Psi$, the specific context makes reference to the particular experimental (quantum) situation, $QP_{\Psi, \mathcal{C}}$, in which the nodes (powers) and their intensive values (potentia) computed through the Born rule  can be exposed through a statistical analysis.

Now that we have the mathematical definition of a PSA, let us go back to the analysis of \emph{pure states}. For simplicity, let us work in $\mathbb{C}^2$. The following analysis can be carried out without difficulties to any dimension. As we defined mathematically (Definition \ref{pure}), a pure state is a unit vector $v\in\mathbb{C}^2$ or in terms of density matrices, it is a $2\times 2$ hermitian matrix
$\rho$ of the form $|v\rangle\langle v|$,
\[
\rho=\begin{pmatrix}
|a|^2& a\overline{b}\\\overline{a}b&|b|^2
\end{pmatrix},\quad v=(a,b),\quad
|a|^2+|b|^2=1.
\]
Notice that 
\[
\rho^2=|v\rangle\langle v|v\rangle\langle v|=|v\rangle\langle v|=\rho
\]
and
\[
\mbox{Tr}(\rho^2)=
|a|^4+|a|^2|b|^2+
|a|^2|b|^2+|b|^4=
|a|^2(|a|^2+|b|^2)+
(|a|^2+|b|^2)|b|^2=1.
\]
Let us translate this representation to our formalism. First, the graph 
of immanent powers $\mathcal{G}$
in $\mathbb{C}^2$ can be pictured as follows,
\begin{center}
\includegraphics[width=19em]{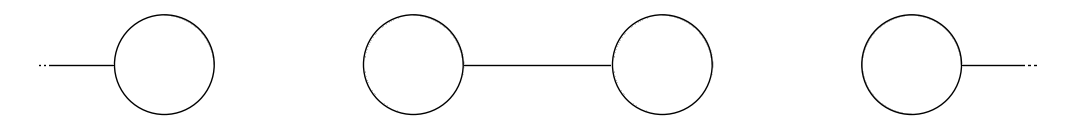}
{\small \fig: Graph of immanent powers in dimension two.}
\end{center}
The previous graph continues to the left and right indefinitely.

Let us choose the basis $v_1=(1,0)$ and $v_2=(0,1)$. This is represented as choosing a maximal context, that is, 
a \emph{complete set of commuting observables}
$\mathcal{C}=\{|v_1\rangle\langle v_1|,|v_2\rangle\langle v_2|\}$,
\begin{center}
\includegraphics[width=19em]{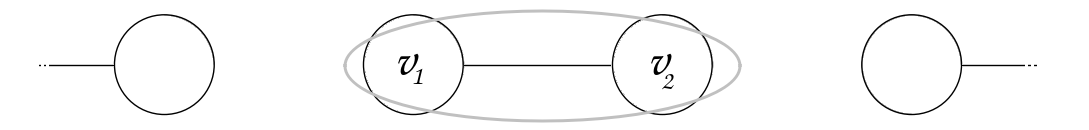}
{\small \fig: Graph with the context $\mathcal{C}$ pointed out.}
\end{center}
Now, we define the PSA $\Psi:\mathcal{G}\to[0,1]$ by using the Born rule.
In this example, 
\[
\mbox{Tr}(\rho\cdot |v_1\rangle\langle v_1|)=|a|^2,\quad
\mbox{Tr}(\rho\cdot |v_2\rangle\langle v_2|)=|b|^2.
\]
We picture the restriction of $\Psi$ to the context $\mathcal{C}$ as
\begin{center}
\includegraphics[width=7em]{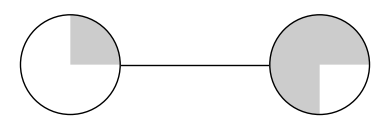}
{\small \fig: The context $\mathcal{C}$ with the assigned potentia to each power.}
\end{center}
But of course, we can also choose another basis. In fact, we can choose the orthonormal basis given by $v=(a,b)$ and $w=(-b,a)$, $a,b\in\mathbb{R}$. Over this context, the representation of $\Psi$ is rather easy,
\begin{center}
\includegraphics[width=7em]{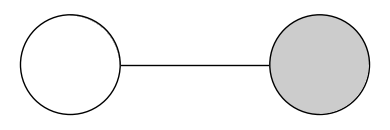}
{\small \fig: Another context showing a pure state.}
\end{center}
As we mentioned above, it is only this particular basis which contains a clear physical operational counterpart relating `the state' to the `{\it certain prediction} of a measurement outcome'. Indeed, since ---following the empiricist-positivist agenda--- it is only the actual {\it and} observable which can find a place in some model of the theory, certain knowledge becomes restricted to actual observable values.\footnote{Something that in Dirac's formulation also links vector kets to the definition of {\it states}.} On the contrary, in the logos approach, none of these powers is problematic since all of them provide objective intensive knowledge of the state of affairs described by QM \cite{deRonde16a} independently of the choice of the context. 

Through the use of graphs we can now visualize very easily the fundamental equivocity present within the different ---both contextual and non-contextual--- definition(s) of {\it pure state} already addressed in \cite{deRondeMassri19b}. As we discussed above, while the mathematical definition makes reference to an abstract context-independent vector (i.e., an invariant with respect to the trace), the operational counterpart is clearly context-dependent and restricts itself to a particular basis (i.e., the basis in which there exists one power with potentia equal to 1). The following graph (figure 6) shows the simultaneous reference of the notion of {\it pure state}, first, to an abstract vector in Hilbert space, second, to a vector represented in a specific basis, and third, to a single {\it eigenvector} whose {\it eigenvalue} is 1. Clearly, each of these mathematical elements possesses not only a distinct mathematical definition, they also codify a completely different type of information. In the logos approach, through the use of graphs we understand visually the confusion present in the orthodox literature according to which a {\it pure state} makes reference, at the same time, firstly, to the whole PSA; secondly, to the single filled node; and thirdly, also to any maximal context containing this node. The scrambling of these three distinct levels of mathematical representation is clearly problematic since we have explicitly shown in the logos approach that there is obviously a difference between considering the whole graph (i.e., a PSA $\Psi$), a particular section of the graph (i.e., a quantum perspective $QP_{\Psi, \mathcal{C}}$), and a particular node of the graph (i.e., an intensive power $P_i$). 
\begin{center}
\includegraphics[width=12em]{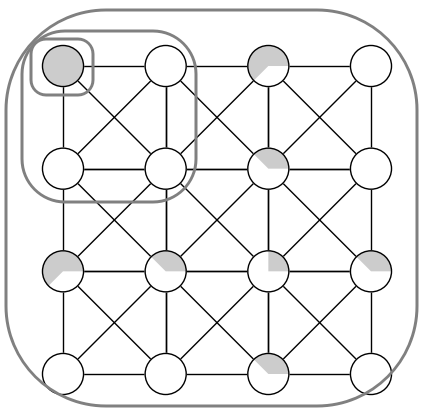}
{\small \fig: The orthodox referential equivocity present within the definition of pure state(s).}
\end{center}
To summarize, while in the topos approach  it is necessary to redefine the formalism in order to incorporate mixed states (matrices of rank grater than 1), in the logos approach all matrices can be  considered on equal footing ---as PSAs or quantum perspectives--- right from the start, without the need to introduce the distinction between pure sates and mixtures. All matrices are treated in a natural and simple way without having to abandon the orthodox formalism or adding anything ``by hand'' like `particles' or `collapses'. While the topos approach attempts to restore the orthodox ignorance interpretation of mixtures, the logos approach avoids any reference to purity and mixtures right from the start. While the topos approach remains tied ---in a Bohrian fashion--- to the contextual reference of a binary (actual) state of affairs, the logos restores the possibility to think of QM as a theory capable of producing an invariant-objective (basis and subject-independent) representation of a (potential) state of affairs in a consistent manner. Finally, while the topos is extremely complicated, both formally and conceptually, the logos approach presents a simple introduction to QM through the theory of graphs as well as an intuitive conceptual explanation of what is really going on according to QM through the notion of {\it intensive power} (see for a detailed discussion of the intuitive content of the logos approach \cite{deRondeMassri18a, deRondeMassri18b, deRondeMassri19a}).

\section{An Intensive-Invariant Representation of Quantum Reality}

In a truly Spinozian spirit, we might say that, in the logos approach ---contrary to orthodoxy--- there are no states or bases which can be considered as more important or fundamental than others; all  bases in QM are {\it as} important. From a physical standpoint which recognizes that invariance is a necessary condition for any consistent physical representation there can be no preferred bases. Our representation in terms of {\it graphs} makes explicit the invariant nature of all reference frames which, in turn, provide access to all the relations present within each specific potential state of affairs. An intensity of a node (a power) equal to $0.5$ and an intensity equal to 1, both provide the same complete accurate type of certain intensive knowledge. The so called {\it actual} properties become just a particular case of {\it potential} or {\it indefinite} properties, just like probability equal to 1 is a particular value of probability not essentially different from probability equal to $0.5$ or $0.77$. Actual (or certain) properties are just a particular case of potential properties, those with potentia =1. Consequently, also from a purely mathematical perspective, a (pure) state $\rho = \rho^2$ and a (impure) state $\rho \neq \rho^2$, are regarded as physically equivalent. Both states provide particular graphs with different tables of powers and potentia, the fact that in the first case there exists a power which has a potentia = 1 is completely irrelevant both from a physical or mathematical perspective which considers an intensive form of quantufication.   

Carlo Rovelli has recently argued \cite{Rovelli18} that Sch\"odinger introduced ``the notion of `wave function' $\psi$, soon to be evolved into the notion of `quantum state' $\psi$, endowing it with heavy ontological weight. This conceptual step was wrong, and dramatically misleading. We are still paying the price for the confusion it has generated.'' Indeed, as we have discussed above, the deep confusion and misunderstanding comes, partly, from the equivocity introduced by the orthodox notation which is unable to account for the different levels of mathematical representation present within the formalism of the theory. But this equivocity has been created by the inadequate idea according to which `QM obviously talks about systems'. Thus, the problem is not that $\psi$ is understood in ontological terms, the problem is that its understanding has been dogmatically restricted to the classical atomist representation in terms of space-time systems. In the logos approach we have provided not only a notation which makes explicit the distinction between the different mathematical levels of representation, we have also provided a conceptual framework in which the mathematical formalism finds a natural connection to operationally well defined invariant physical concepts. While the non-contextual aspect of abstract vectors is described in terms of a PSA, the contextual nature of quantum superpositions is clearly stressed through the reference to the notion of `quantum perspective'.  In this respect, an important aspect of our logos approach is that all these new (non-classical) notions possess a physical operational-invariant counterpart. Just like Einstein required, all these ---newly introduced--- physical concepts contain the operational conditions allowing to discover whether or not they are fulfilled in an actual case. In this respect, in order to restore the complete operational equivalence between all projection operators it might be recommendable to return to Heisenberg's original formulation of QM in which all matrices, independently of their rank, stand on equal theoretical and operational footing. It is within this matrix formulation ---in which both notions of purity and mixture becomes untenable--- that the logos approach might find its natural extension. Something we leave for a future work.

\section*{Conclusion}

In this paper we have discussed the untenability of the notion of {\it pure state} in the orthodox formalism of QM. Through the aid of graphs we have shown the equivocity present within the different definitions confused and scrambled in the present literature. We have also shown that through the application of an intensive analysis it is possible to restore an objective theoretical representation of QM. In this new scheme it becomes explicit why the distinction between {\it pure state} and {\it mixed state} is completely irrelevant both from a mathematical and a physical perspectives.

\section*{Acknowledgements} 

This work was partially supported by the following grants: FWO project G.0405.08 and FWO-research community W0.030.06. CONICET RES. 4541-12, the Project UNAJ 80020170100058UJ and PIO-CONICET-UNAJ 15520150100008CO ``Quantum Superpositions in Quantum Information Processing''.

\end{document}